# Matrix-free numerical continuation of the Ikeda map in passive Kerr resonators

**YUANDI XU,[1,†] PENGXIANG WANG,[1,†] JIANXING PAN,[2] HAISU ZHANG,[3,*] AND GANG XU[1,*]**

[1]*School of Optical and Electronic Information, Huazhong University of Science and Technology, Wuhan, 430074, China*
[2]*Hangzhou International Innovation Institute, Beihang University, Hangzhou, 311115, China*
[3]*School of Physics and Mechatronic Engineering, Guizhou Minzu University, Guiyang, 550025, China*
[†]*These authors contributed equally to this work.*
*[*hszhang@phy.ecnu.edu.cn, gang_xu@hust.edu.cn]*

**We introduce a matrix-free numerical continuation framework for the Ikeda map of coherently driven passive Kerr resonators. The framework traces solution branches and evaluates their linear stability without explicitly assembling the full Jacobian matrices, using a unified formulation for states that recur after one or multiple round trips. We resolve stable and unstable branches of modulation-instability patterns and temporal cavity solitons, and uncover period-doubled homogeneous states and associated multi-round-trip dynamics that are absent from the standard Lugiato–Lefever equation (LLE). For period-1 states, the Ikeda map and the LLE show close agreement in the high-finesse regime, while their branch-existence intervals increasingly deviate as the finesse decreases. These results demonstrate that direct continuation of the unaveraged round-trip map can capture nonlinear states and multi-round-trip dynamics beyond the mean-field description, providing a systematic framework for bifurcation and stability analysis of Kerr resonators without relying on the averaged LLE.**

Coherently driven passive Kerr resonators have attracted sustained interest in nonlinear optics as platforms for optical information processing [1], precision metrology [2], spectroscopy [3], and optical communications [4]. In these resonators, dispersion and Kerr nonlinearity shape the intracavity field, while coherent driving compensates for cavity losses. Their interplay gives rise to rich nonlinear behavior, including pattern formation [5], temporal cavity solitons, and the associated optical frequency combs [6–8]. The round-trip dynamics of these resonators can be described by the Ikeda map [7–9]. It combines a nonlinear Schrödinger equation (NLSE) describing intracavity propagation with a discrete boundary condition relating the fields on successive round trips.

For theoretical analysis and numerical modeling, the Lugiato–Lefever equation (LLE) [10,11] offers a simplified description of Kerr resonator dynamics. The LLE can be derived from the Ikeda map under the mean-field approximation, where discrete pump injection and coupling losses are averaged over the cavity length [8]. Over the last four decades, the LLE framework has been extensively adopted to investigate Kerr cavity dynamics and has successfully reproduced experimentally observed temporal cavity solitons, as well as the associated Kerr frequency combs [1,12–14]. Despite its widespread utility, the standard LLE is only quantitatively reliable for high-finesse cavities near resonance [8,15]. This model breaks down when distinct nonlinear cavity resonances overlap or when the field changes substantially between successive round trips [8,16]. In such regimes, the Ikeda map preserves the discrete round-trip field evolution and thus captures physical effects inaccessible to conventional mean-field theories. Examples include overlapping nonlinear resonances, round-trip period doubling (PD) [17,18], and super cavity solitons [19,20]. To address this limitation, alternative modeling approaches have been proposed for these operating regimes, such as multi-resonant extensions of the LLE [16] and time-delay models [21]. Nevertheless, directly performing systematic continuation of solution branches and bifurcation analysis within the native Ikeda map framework remains technically challenging.

In this Letter, we overcome this obstacle by performing numerical continuation directly on the Ikeda map using a matrix-free Newton–Krylov corrector [22] and pseudo-arclength continuation. Standard pseudo-arclength continuation [23] has been used to study the LLE and the low-dimensional dispersionless Ikeda map [24]. For the dispersive Ikeda map, however, fast-time discretization yields a large fixed-point system whose Jacobian accounts for the full round-trip propagation. Explicit construction and storage of this matrix can be costly. Our matrix-free framework avoids explicit Jacobian construction and storage while tracking stable and unstable branches of period-1 (P1) and higher-period states and evaluating their Floquet stability. We compare P1 branches and cavity-soliton properties with LLE predictions, quantifying improved agreement at higher cavity finesse. We also resolve round-trip period-doubled states beyond conventional mean-field descriptions. Operating directly on the evolution map, our framework is in principle adaptable to other discrete driven-dissipative systems without deriving auxiliary models.

Our starting point is a coherently driven Kerr ring cavity, simplified as a waveguide segment closed by a coupler (see schematic illustration in Fig. 1). The intracavity field evolves continuously along the resonator and undergoes a discrete update at the coupler through pump injection and partial out-coupling. The resulting round-trip dynamics are described by the Ikeda map [7–9]. We write it in the dimensionless variables used for the LLE [25] (see Section 1 of Supplement 1 for normalization details):

$$\frac{\partial E^m(z,\tau)}{\partial z} = -\frac{\alpha_{\rm rt}}{2}E^m(z,\tau) - i\alpha\eta\frac{\partial^2 E^m(z,\tau)}{\partial \tau^2} + i\alpha|E^m(z,\tau)|^2E^m(z,\tau), \quad (1)$$

$$E^{m+1}(0,\tau) = \alpha S + \sqrt{1-\theta}\,e^{-i\alpha\Delta}E^m(1,\tau). \quad (2)$$

Here, the superscript $m$ is the round-trip index, and $E^m(z,\tau)$ is the intracavity field during round trip $m$. The coordinate $z \in [0,1]$ runs from immediately after the coupler to just before the subsequent boundary update, while $\tau$ corresponds to the retarded fast time in a frame moving at the group velocity evaluated at the pump frequency. The parameters $S$ and $\Delta$ denote the pump amplitude and cavity detuning, respectively, while $\eta = -1$ specifies the anomalous-dispersion regime considered here. The parameter $\alpha_{\rm rt} = \alpha_i L$ is the intrinsic round-trip power-loss exponent, where $\alpha_i$ is the distributed intrinsic power-attenuation coefficient and $L$ is the cavity length. The parameter $\theta$ is the power coupling coefficient of the coupler. Together, they define $\alpha = (\theta + \alpha_{\rm rt})/2$ and the round-trip field-retention factor $q = \sqrt{1-\theta}\exp(-\alpha_{\rm rt}/2)$. The cavity finesse $\mathcal{F} = \pi/\{2\arcsin[(1-q)/(2\sqrt{q})]\}$, defined as the ratio of the free spectral range to the resonance linewidth, characterizes the overall cavity loss. The external-coupling fraction $\chi = -\ln(1-\theta)/[-\ln(1-\theta) + \alpha_{\rm rt}]$ quantifies the contribution of external coupling to the total logarithmic round-trip power loss. We set $\chi = 0.5$ throughout and vary $\mathcal{F}$ to assess deviations from the mean-field limit.

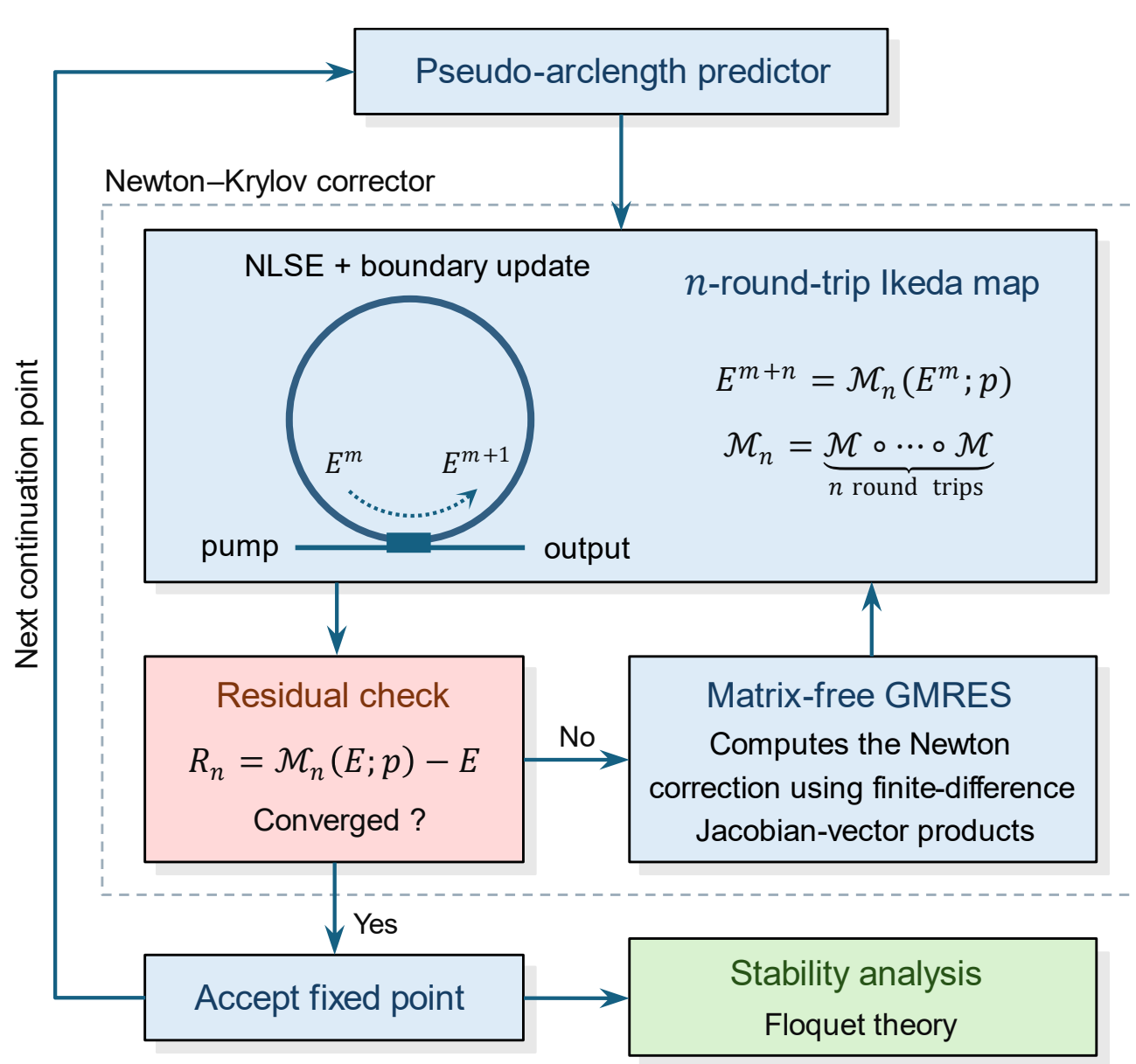

**Fig. 1.** Workflow for matrix-free numerical continuation of the Ikeda-map fixed points, followed by Floquet stability analysis.

To trace the solution branches of the Ikeda map using numerical continuation, we formulate its round-trip dynamics as a fixed-point problem. Let $\mathcal{M}(E;\mathrm{p})$ denote the one-round-trip map acting on the post-coupler intracavity field $E$, with the detuning $\Delta$ used as the continuation parameter $p$. Each application of $\mathcal{M}$ propagates the field from $z = 0$ to $z = 1$ using the split-step Fourier method [26] and then applies the coupler boundary condition. The $n$-fold composition of $\mathcal{M}$ defines the $n$-round-trip map $\mathcal{M}_n$. Its fixed points satisfy

$$R_n(E;p) \equiv \mathcal{M}_n(E;p) - E = 0. \quad (3)$$

The case $n = 1$ corresponds to P1 states.

We then solve the fixed-point problem within a matrix-free Newton–Krylov continuation framework [27,28]. This approach has been used to continue invariant solutions in fluid and thermoacoustic systems [22,27] and, more recently, branches of periodic orbits in the LLE [29] and coupled LLEs [30]. Our implementation, summarized in Fig. 1, combines a matrix-free Newton–Krylov corrector with a pseudo-arclength predictor. In this formulation, the unknowns are collected into a solution vector comprising the discretized real and imaginary parts of $E$ together with the continuation parameter $p$. Starting from an accepted solution, the predictor advances along the local branch tangent to provide an initial estimate for the Newton–Krylov corrector.

At each Newton iteration of the corrector, the fixed-point residual and the pseudo-arclength constraint are linearized to form a bordered system for the Newton corrections $(\delta E, \delta p)$. Let $J = D_E R_n(E;p)$ denote the Jacobian of the fixed-point residual with respect to the field, where $D_E$ denotes differentiation with respect to $E$ at fixed $p$. For a field perturbation $v$, $Jv$ represents the first-order change in the residual induced by perturbing $E$ along $v$. This product can be approximated by the first-order finite difference [28], $Jv \approx [R_n(E+\varepsilon v;p) - R_n(E;p)]/\varepsilon$, where $\varepsilon$ is a small finite-difference step. The residual derivative $D_p R_n(E;p)$ and the linearized pseudo-arclength constraint provide the remaining column and row of the bordered operator, respectively. The generalized minimal residual (GMRES) method [31] then builds a Krylov basis through repeated evaluations of bordered Jacobian-vector products and obtains the Newton correction from the associated reduced least-squares problem, without assembling the bordered Jacobian matrix. Each Newton correction updates both $E$ and $p$, after which the $n$-round-trip map is reevaluated. The correction cycle continues until the fixed-point and pseudo-arclength residuals satisfy the prescribed tolerances. The converged solution is then accepted, and its updated tangent supplies the predictor for the next continuation point.

Once a continuation point has been accepted, we assess its linear stability using the linearized $n$-round-trip operator $D_E\mathcal{M}_n(E;p)$. This operator maps an infinitesimal perturbation of the post-coupler field to the resulting perturbation after $n$ round trips. For a perturbation $v$, the product $D_E\mathcal{M}_n(E;p)v$ is evaluated by propagating $v$ through the linearized intracavity equation and the corresponding coupler boundary update over $n$ round trips. An Arnoldi iteration [32] computes the dominant eigenvalues $\rho$ through repeated evaluations of these operator-vector products, without explicitly forming the matrix representation of $D_E\mathcal{M}_n(E;p)$. These eigenvalues are the dominant Floquet multipliers. For nonuniform states, continuous translational symmetry in fast time produces a neutral multiplier at $\rho = 1$. Excluding this multiplier, linear stability requires all remaining multipliers to lie strictly inside the unit circle [23].

We first apply the continuation framework to homogeneous steady states (HSS), which can undergo period-doubling bifurcations in Ikeda-type systems [17,18]. The HSS case provides a convenient test of the framework because the dispersive term vanishes in both the fixed-point equation and the linearized problem for homogeneous perturbations. Figure 2(a) compares the continued P1 and period-2 (P2) branches of the Ikeda map with the HSS response of the LLE. In the Ikeda map, the $2\pi$-periodicity of the boundary phase gives rise to a sequence of tilted resonances, whereas the standard LLE describes a single resonance [7].

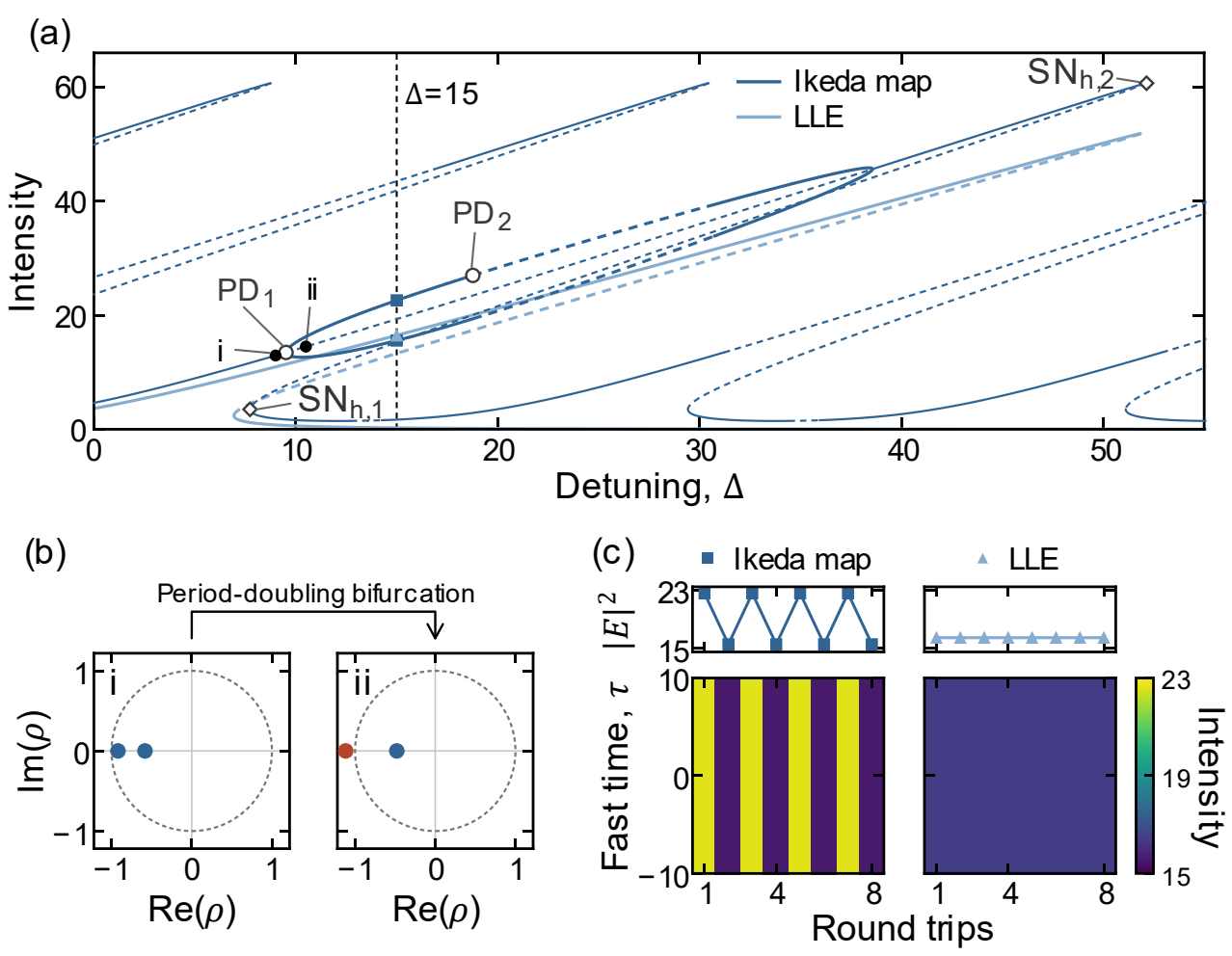

**Fig. 2.** Comparison of homogeneous states in the Ikeda map ($\mathcal{F} = 10$, dark) and the LLE (light) at $S = 7.2$. (a) Intensity versus detuning, including the Ikeda-map P1 and P2 branches. Solid (dashed) curves are stable (unstable) to homogeneous perturbations. Diamonds and circles mark saddle-node (SN) and period-doubling (PD) bifurcations, respectively. (b) P1 Floquet multipliers at (i) $\Delta = 9$ and (ii) $\Delta = 10.5$, before and after $PD_1$. Red dots indicate unstable multipliers with $|\rho| > 1$. (c) Round-trip intensity evolution of the Ikeda-map P2 state and the upper LLE state at $\Delta = 15$.

On the lower-detuning side of one such resonance, the P1 branch undergoes a period-doubling bifurcation at $\Delta = 9.5$ ($PD_1$), giving rise to the P2 branch. The P2 branch undergoes a second period-doubling bifurcation at $\Delta = 18.8$ ($PD_2$), giving rise to a period-4 branch (see Section 2 of Supplement 1 for the subsequent period-doubling cascade). Figure 2(b) compares the P1 Floquet spectra for homogeneous perturbations before (i) and after (ii) $PD_1$. Both multipliers lie inside the unit circle at $\Delta = 9$, whereas one real multiplier lies below $-1$ at $\Delta = 10.5$, marking the loss of stability to homogeneous perturbations. At $\Delta = 15$, the P2 intensity alternates between two values over successive round trips, while the upper LLE HSS remains stationary [Fig. 2(c)]. The Ikeda map thus reveals period-doubled round-trip dynamics hidden by mean-field averaging.

We next extend the comparison to P1 modulation-instability (MI) patterns and cavity solitons (CSs) at $S = 2.8$, with $\mathcal{F} = 20$ for the Ikeda map [Fig. 3(a)]. Unlike the homogeneous state in Fig. 2, these solutions vary with fast time and retain the dispersive term, thereby providing a more stringent test of the Ikeda-map continuation framework. In both models, an MI branch bifurcates from the homogeneous state, with its modulation frequency set to the most unstable one predicted by the LLE at threshold [25]. The CS branch folds at a high-detuning saddle-node (SN) and returns toward lower peak intensity, defining the high-detuning boundary of the stationary CS existence interval. With this cavity finesse, the corresponding Ikeda-map and LLE branches agree closely but do not coincide exactly. At $\Delta = 9.5$, the upper-branch CSs in the two models exhibit similar single-pulse intensity profiles, with slightly different peak intensities [Fig. 3(b)].

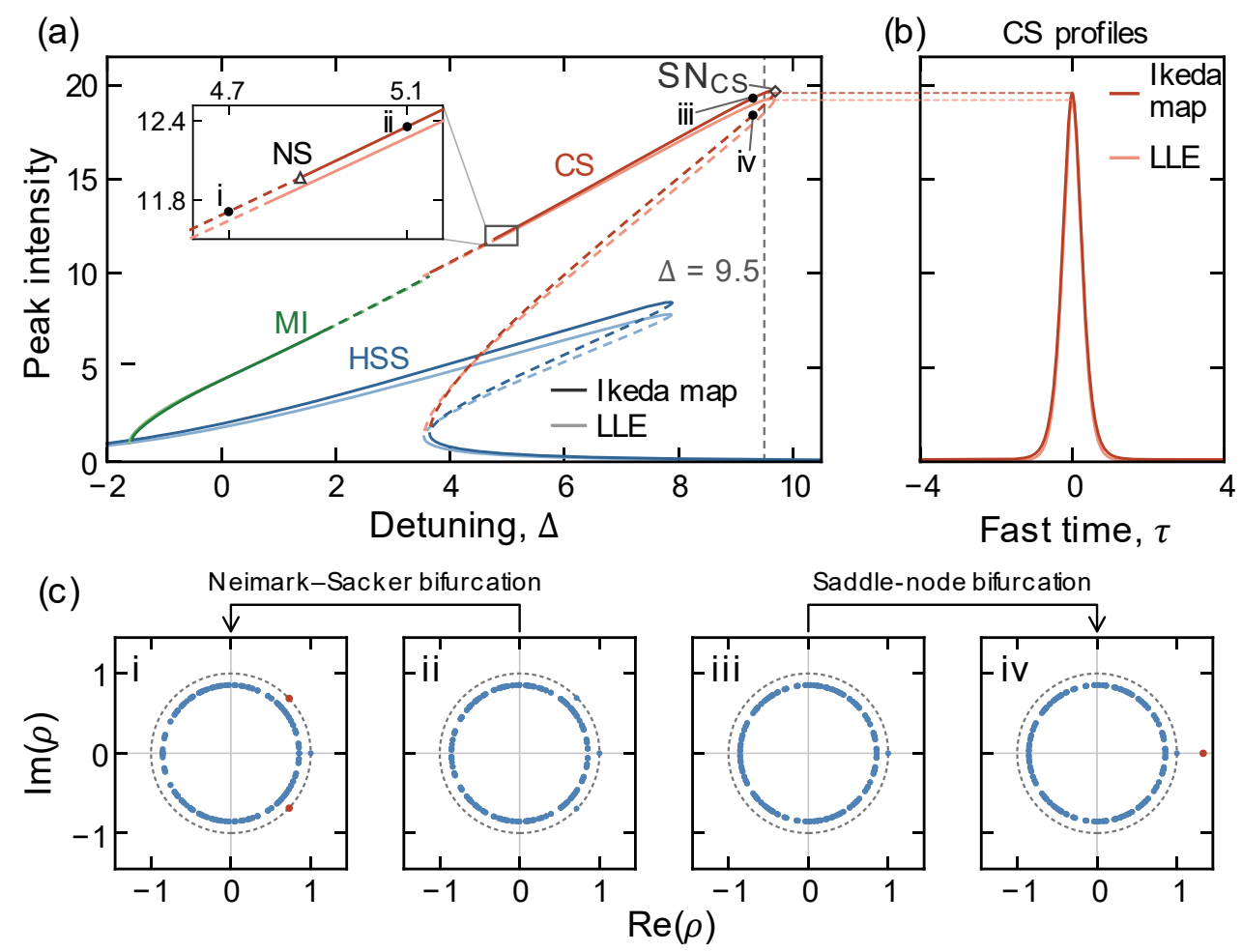

**Fig. 3.** (a) HSS (blue), MI (green), and CS (red) branches in the LLE (light) and Ikeda map (dark) at $S = 2.8$ and $\mathcal{F} = 20$. Solid (dashed) segments are stable (unstable). HSS stability is assessed against homogeneous perturbations. Diamonds and triangles mark saddle-node (SN) and Neimark–Sacker (NS) bifurcations, respectively. (b) CS profiles at $\Delta = 9.5$. (c) Ikeda-map Floquet multipliers near NS at (i) $\Delta = 4.7$ and (ii) 5.1, and near SN at (iii, iv) $\Delta = 9.3$. Red dots indicate unstable multipliers with $|\rho| > 1$.

We then assess the linear stability of selected Ikeda-map CS states using their one-round-trip Floquet spectra [Fig. 3(c)]. The spectra retain the translational neutral multiplier at $\rho = 1$, reflecting invariance under fast-time shifts. As detuning decreases along the upper branch from $\Delta = 5.1$ (ii) to 4.7 (i), a complex-conjugate multiplier pair crosses outward through the unit circle. The upper-branch CS thus loses stability through a Neimark–Sacker (NS) instability, the discrete-time counterpart of a Hopf instability [23]. Figure 3(c)(iii, iv) compares the upper- and returning-branch states at $\Delta = 9.3$, near the high-detuning saddle-node. The upper-branch state (iii) is linearly stable, whereas the returning-branch state (iv) has a real multiplier $\rho > 1$. This multiplier indicates a nonoscillatory instability, distinct from the complex-pair NS instability and period doubling at $\rho = -1$ [23]. These stability classifications apply to perturbations within the corresponding periodic fast-time window. Continuation therefore recovers unstable CS states that cannot be obtained as long-time attractors through direct time integration.

In order to gain more physical insights, we present Fig. 4, which extends the comparison in Fig. 3 across a range of cavity finesse values at $S = 2.8$. Figure 4(a) compares the continued HSS and CS branches of the Ikeda map at $\mathcal{F} = 5, 10$ and 100 (panels i–iii, respectively) with the LLE reference in the lower-right panel. At low finesse, the CS branches of the Ikeda map and the LLE differ in shape, existence range, and stability. These differences are particularly pronounced at $\mathcal{F} = 5$, where the stable detuning interval of the P1 CS branch is substantially narrower than predicted by the LLE. As $\mathcal{F}$ increases, these differences diminish and the Ikeda-map CS branches approach the LLE prediction. We quantify the agreement between the P1 CS existence intervals (upper-branch intervals) in detuning using the Jaccard overlap [33]. It is defined as the ratio of the intersection length to the union length of these detuning intervals. This detuning-interval overlap increases monotonically toward unity with increasing $\mathcal{F}$ [Fig. 4(b)].

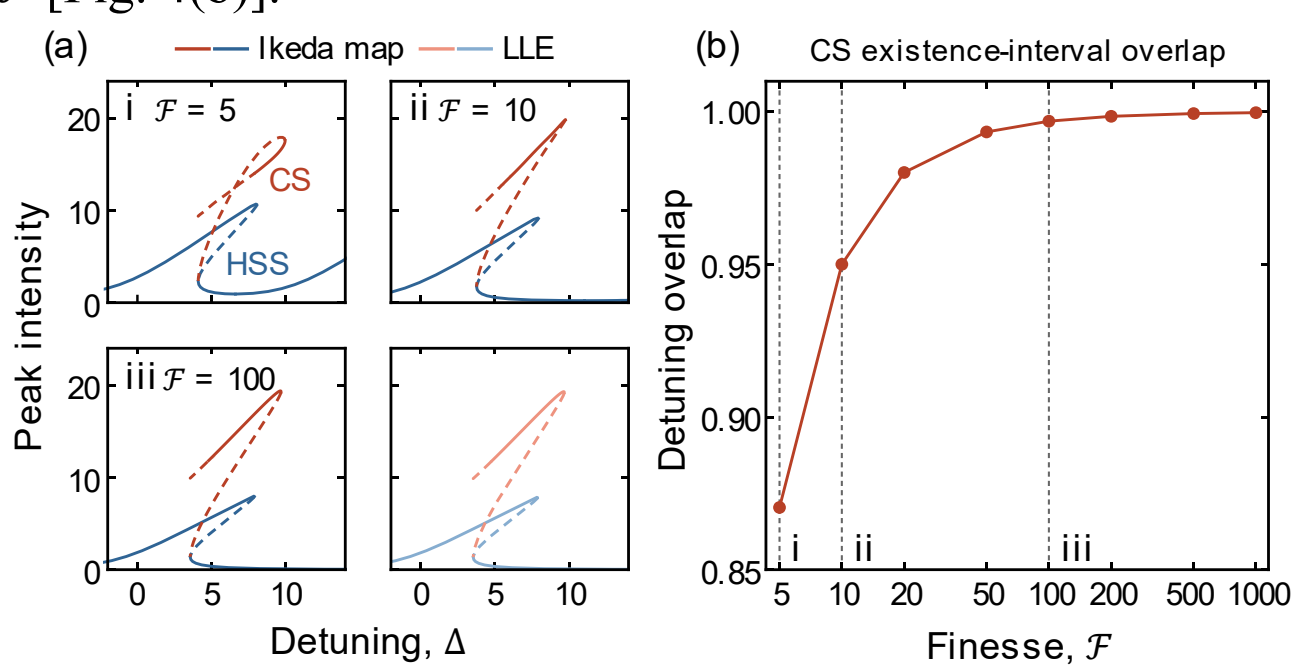

**Fig. 4.** Comparison of the Ikeda map and LLE at $S = 2.8$. (a) HSS (blue) and CS (red) branches in the Ikeda map (dark) at (i–iii) $\mathcal{F} = 5, 10, 100$, respectively. The LLE reference (light) is shown in the lower-right panel. Solid (dashed) segments represent stable (unstable) regimes. HSS stability is assessed against homogeneous perturbations. (b) Detuning-interval overlap of the CS existence ranges versus finesse.

In summary, we have developed a matrix-free Newton–Krylov pseudo-arclength continuation framework for the Ikeda map of coherently driven passive Kerr cavities. By evaluating Jacobian-vector products without explicitly assembling the full Jacobian matrices, the framework efficiently traces both P1 and higher-period solution branches and determines their Floquet stability within a unified formulation. Applied to HSS, MI, and CS states, the method resolves their stable and unstable branches and reveals multi-round-trip dynamics, including period-doubled states, that are absent from the standard LLE description. More importantly, because the continuation is performed directly on the discrete round-trip map, the framework does not rely on the mean-field or good-cavity approximations and can naturally accommodate low-finesse regimes and systems with complex propagation and boundary conditions.

Beyond passive Kerr ring resonators, this formulation is intrinsically compatible with complex propagation dynamics and boundary conditions, and can therefore be generalized to resonators or even laser cavities for which a reliable mean-field model is difficult or impossible to derive. For example, in complex fiber-laser configurations, by directly treating the discrete round-trip map itself as the fundamental dynamical object, this method provides a general route to continuation, bifurcation, and stability analysis. Therefore, it sheds new light on multi-round-trip dynamics and nonlinear states that are hidden by mean-field approximations, offering a broadly applicable framework for driven-dissipative systems with complex cavity architectures.

**Funding.** The National Natural Science Foundation of China (62275097); Zhejiang Provincial Postdoctoral Research Excellence Funding Program (ZJ2025146); China Postdoctoral Science Foundation (2025M784278).

**Disclosures.** The authors declare no conflicts of interest.

**Data availability.** Data underlying the results presented in this paper are not publicly available at this time but may be obtained from the authors upon reasonable request.

## FULL REFERENCES

# Matrix-free numerical continuation of the Ikeda map in passive Kerr resonators: supplemental document

## 1. Normalization of the Ikeda map

The main text employs a dimensionless form of the Ikeda map. For completeness, we provide here the dimensional starting equations and the normalization procedure leading to the dimensionless model. The dimensional Ikeda map consists of the intracavity propagation equation and the coupler boundary condition [1–3]:

$$\frac{\partial E^m(z,\tau)}{\partial z} = -\frac{\alpha_i}{2}E^m(z,\tau) - i\frac{\beta_2}{2}\frac{\partial^2 E^m(z,\tau)}{\partial \tau^2} + i\gamma|E^m(z,\tau)|^2 E^m(z,\tau), \tag{S1}$$

$$E^{m+1}(0,\tau) = \sqrt{\theta}E_{\text{in}} + \sqrt{1-\theta}e^{-i\delta_0}E^m(L,\tau). \tag{S2}$$

Here, $E^m(z,\tau)$ is the dimensional intracavity field envelope during round trip $m$. The coordinate $z$ measures propagation along a cavity of length $L$. The retarded fast time $\tau$ is defined in a reference frame moving at the group velocity. In Eq. (S1), $\alpha_i$ is the intrinsic power-attenuation coefficient, $\beta_2$ is the second-order group velocity dispersion coefficient, and $\gamma$ is the Kerr nonlinear coefficient. In Eq. (S2), $E_{\text{in}}$ denotes the amplitude of the incident pump field, and $\theta$ is the coupler power coupling coefficient. The parameter $\delta_0$ denotes the round-trip phase detuning relative to a reference cavity resonance.

To obtain the dimensionless form used in the main text, we rescale the field and coordinates following the standard Lugiato–Lefever equation (LLE) normalization [4], with the normalized parameters and variables defined by

$$\begin{gathered}\alpha_{\text{rt}} = \alpha_i L, \qquad \alpha = \frac{\theta+\alpha_{\text{rt}}}{2}, \qquad \Delta = \frac{\delta_0}{\alpha}, \qquad S = E_{\text{in}}\sqrt{\frac{\theta\gamma L}{\alpha^3}}, \\ E^m \to E^m\sqrt{\frac{\gamma L}{\alpha}}, \qquad \tau \to \tau\sqrt{\frac{2\alpha}{|\beta_2|L}}, \qquad z \to \frac{z}{L}, \qquad \eta = \text{sgn}(\beta_2)\,.\end{gathered} \tag{S3}$$

Under this normalization, Eqs. (S1) and (S2) become

$$\frac{\partial E^m}{\partial z} = -\frac{\alpha_{\text{rt}}}{2}E^m - i\alpha\eta\frac{\partial^2 E^m}{\partial \tau^2} + i\alpha|E^m|^2 E^m, \tag{S4}$$

$$E^{m+1}(0,\tau) = \alpha S + \sqrt{1-\theta}e^{-i\alpha\Delta}E^m(1,\tau). \tag{S5}$$

In these dimensionless equations, $z \in [0,1]$ is the normalized propagation coordinate, with the interval spanning one cavity round trip. The parameter $\alpha$ is a loss scale combining intrinsic round-trip loss and coupling loss. The parameters $\Delta$ and $S$ denote the dimensionless cavity detuning and pump amplitude, respectively. The sign parameter $\eta = \text{sgn}(\beta_2)$ specifies the dispersion regime, with $\eta = -1$ corresponding to the anomalous-dispersion regime considered here.

## 2. Period-doubling cascade of homogeneous states in the Ikeda map

Figure S1 extends the continuation results shown in Fig. 2(a) of the main text to higher-period homogeneous states, thereby documenting the successive period-doubling cascade of the Ikeda map. The calculations are performed at pump amplitude $S = 7.2$, cavity finesse $\mathcal{F} = 10$, and external-coupling fraction $\chi = 0.5$. At $\Delta = 9.5$, the period-1 (P1) branch undergoes a period-doubling (PD) bifurcation $PD_1$, from which the period-2 (P2) branch emerges. The P2 and period-4 (P4) branches subsequently undergo further period-doubling bifurcations at $\Delta = 18.8$ and $21.7$, producing the P4 and period-8 (P8) branches, respectively. A further period-doubling bifurcation is detected on the P8 branch at $\Delta = 22.7$, consistent with the emergence of a period-16 branch; the latter is not continued here. The enlarged view in Fig. S1(b) resolves the closely spaced bifurcations of the higher-period branches.

The post-coupler intracavity intensities in Figs. S1(c)–S1(e) are shown over 16 round trips for representative P2, P4, and P8 states at $\Delta = 18.5$, $21$, and $22.2$, respectively. Their alternation among two, four, and eight distinct values illustrates the corresponding discrete round-trip periods.

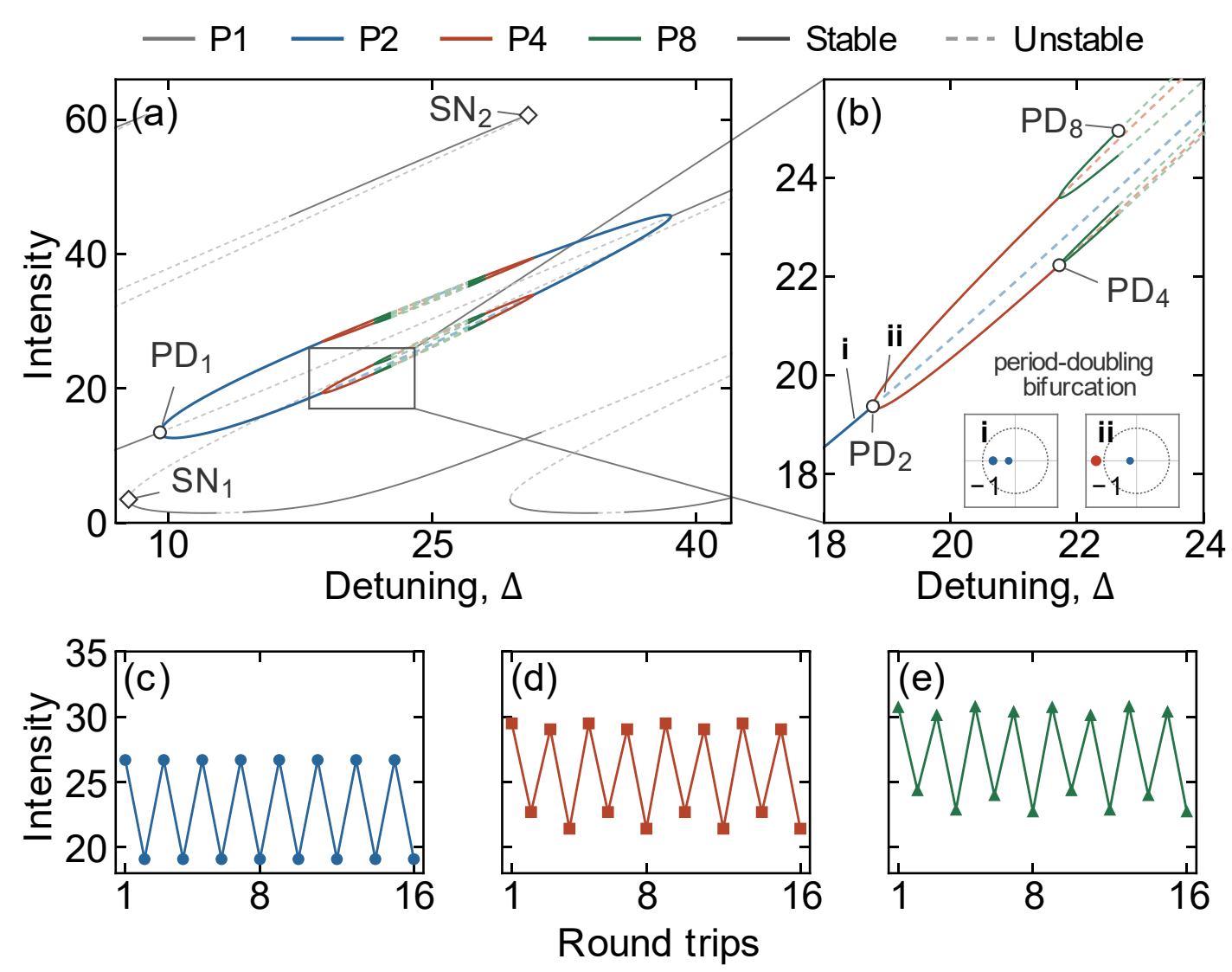

**Fig. S1.** Extended period-doubling cascade of homogeneous states in the Ikeda map at $S = 7.2$, cavity finesse $\mathcal{F} = 10$, and external-coupling fraction $\chi = 0.5$ (see main text for definition). (a) P1 (gray), P2 (blue), P4 (red), and P8 (green) branches. Dark solid (light dashed) curves are stable (unstable) under homogeneous perturbations. (b) Enlargement of the P2–P4–P8 cascade. The multiplier configurations at i and ii schematically illustrate a real Floquet multiplier crossing $\rho = -1$ at $PD_2$. (c)–(e) Intensities over 16 round trips for P2 at $\Delta = 18.5$, P4 at $\Delta = 21$, and P8 at $\Delta = 22.2$, respectively.